\documentclass[english]{revtex4-1}
\usepackage[T1]{fontenc}
\usepackage[latin9]{inputenc}
\usepackage{textcomp}
\usepackage{amsmath}
\usepackage{amssymb}

\makeatletter
\usepackage{babel}

\makeatother

\usepackage{babel}
\begin{document}

\title{No relativistic probability current for any spin}

\author{Scott E. Hoffmann}

\address{School of Mathematics and Physics,~~\\
 The University of Queensland,~~\\
 Brisbane, QLD 4072~~\\
 Australia}
\email{scott.hoffmann@uqconnect.edu.au}

\selectlanguage{english}%
\begin{abstract}
We investigate whether the Newton-Wigner position probability density,
extended from spinless particles to electrons/positrons and particles
of higher spin, can be incorporated as the zero component of a four-component
probability current density that transforms locally as a four-vector
function of the spacetime coordinates. We find that this is not possible,
in all cases.
\end{abstract}
\maketitle

\section{Introduction}

Perhaps the most important result of Dirac's theory of the electron
and positron \cite{Dirac1928} is the construction of the four-component
charge current density operator (as a single particle operator)
\[
J_{\mathrm{D}}^{\mu}(x)=q\frac{m_{e}}{(2\pi)^{3}}\int\frac{d^{3}p_{a}}{\omega_{a}}\sum_{m_{a}=\pm\frac{1}{2}}\int\frac{d^{3}p_{b}}{\omega_{b}}\sum_{m_{b}=\pm\frac{1}{2}}|\,p_{a},m_{a}\,\rangle\,e^{i(p_{a}-p_{b})\cdot x}\,\bar{u}(p_{a},m_{a})\gamma^{\mu}u(p_{b},m_{b})\langle\,p_{b},m_{b}\,|,
\]
with the spinors and gamma matrices as in \cite{Itzykson1980}. It
transforms locally as a four-vector function of the spacetime coordinates
for all Lorentz transformations, $L,$
\begin{equation}
U^{\dagger}(L)\,J_{\mathrm{D}}^{\mu}(x)\,U(L)=L_{\phantom{\mu}\nu}^{\mu}\,J_{\mathrm{D}}^{\nu}(L^{-1}x).\label{eq:1.1}
\end{equation}
It translates in spacetime as
\begin{equation}
J_{\mathrm{D}}^{\mu}(x+a)=U(T(a))\,J_{\mathrm{D}}^{\mu}(x)\,U^{\dagger}(T(a)).\label{eq:1.2}
\end{equation}
It is locally conserved,
\begin{equation}
\partial_{\mu}J_{\mathrm{D}}^{\mu}(x)=0,\label{eq:1.3}
\end{equation}
and as a consequence the charge density is globally conserved in all
frames:
\begin{equation}
\int d^{3}x\,J_{\mathrm{D}}^{0}(x)=q.\label{eq:1.4}
\end{equation}
It contains both momentum- and spin-dependent contributions, as can
be seen using the Gordon decomposition \cite{Itzykson1980}:
\[
\overline{u}(p_{a},m_{a})\gamma^{\mu}u(p_{b},m_{b})=\frac{1}{2m_{e}}(p_{a}+p_{b})^{\mu}\overline{u}(p_{a},m_{a})u(p_{b},m_{b})+\frac{i}{2m_{e}}\overline{u}(p_{a},m_{a})\sigma^{\mu\nu}u(p_{b},m_{b})(p_{a}-p_{b})_{\nu}
\]
Furthermore, the charge density, the zero component of the four-vector,
is negative (positive) semidefinite for the electron (positron).

Note that this result can be used for electrons, by taking $q=-e,$
or positrons, by taking $q=+e.$ The identity
\begin{equation}
\bar{v}(p_{a},m_{a})\gamma^{\mu}v(p_{b},m_{b})=-\bar{u}(p_{a},m_{a})\gamma^{\mu}u(p_{b},m_{b})\label{eq:1.5}
\end{equation}
shows that it is not necessary to use $v$ spinors in this expression.

We know that this form of the current does not entirely represent
the physics. It predicts a g-factor for the electron or positron magnetic
moment of exactly 2. The experimental result is $g_{e}=2+2(1159.65218091\text{\textpm}0.00000026)\times10^{-6}$
\cite{ParticleDataGroup2018}. With quantum field theory, the corrections
can be calculated. However it would be desirable to see a modified
form of the current, taking into account these corrections, hopefully
with the charge density remaining negative (positive) semidefinite.
One way to do this would be to apply a scalar unitary transformation
to the current.

There is a way, consistent with special relativity, of defining a
spatial probability density for an electron or positron. In this paper
we investigate whether this probability density can be taken as the
zero component of a four-component probability current that transforms
like the Dirac current and is conserved. We will find that this is
not possible for the electron or positron, for a spinless particle
or, in fact, for any spin.

If such a probability current were possible, we suppose that a charge
current could be obtained simply by multiplying by the charge. If
this were possible we might have a second way to construct a quantum
electrodynamics, and would have to compare with the standard results.
The impossibility of this construction gives us confidence in the
Dirac current density as (close to) the correct representation of
the physical current.

The spinless case was treated by Rosenstein and Horwitz \cite{Rosenstein1985},
where they claim to have constructed such a relativistic probability
current. We find that they are in error.

The organization of this paper is as follows. First, in Section II,
we review results on relativistic probability amplitudes and the construction
of a total position probability density (summed over spins) for the
electron or positron. In Section III we derive the fundamental commutation
relations between the boost generators and a four-vector. We also
find the representation of the boost generators as differential operators
acting on momentum-spin component wavefunctions. In Section IV we
construct the spatial components of what is possibly a four-vector
probability current density for the electron/positron. Then we show
that the second set of commutators from Section II is not satisfied
for these four components. In Section V we briefly consider the spinless
case and show, again, that a four-vector probability current density
cannot be constructed. This requires examination of the result of
Rosenstein and Horwitz to show where they are in error. The case of
general spin is then considered, with the same conclusion of no possible
four-vector probability current density. Conclusions follow in Section
VI.

\section{Relativistic probability amplitudes and the position probability
density}

We begin with the improper basis vectors, $|\,p,m\,\rangle\ (m=\pm1/2),$
that carry the unitary, irreducible representations of the Poincaré
group \cite{Wigner1939} for free electrons and positrons. They are
eigenvectors of four-momentum with eigenvalue components $p^{\mu}=(\omega,\boldsymbol{p})^{\mu}.$
We deal only with positive energies $p^{0}=\omega=\sqrt{\boldsymbol{p}^{2}+m_{e}^{2}}.$
The spin label, $m,$ carries the representation of rotations in the
rest frame. A charge label is to be understood, the only action necessary
to distinguish positrons from electrons. We choose to give these basis
vectors the covariant normalization
\begin{equation}
\langle\,p_{1},m_{1}\,|\,p_{2},m_{2}\,\rangle=\omega_{1}\delta^{3}(\boldsymbol{p}_{1}-\boldsymbol{p}_{2}).\label{eq:2.1}
\end{equation}

If we contruct a wavepacket state vector, a superposition of the basis
vectors, normalized to unity, as
\begin{equation}
|\,\psi\,\rangle=\int\frac{d^{3}p}{\sqrt{\omega}}\sum_{m=\pm\frac{1}{2}}|\,p,m\,\rangle\Psi_{m}(p),\label{eq:2.2}
\end{equation}
then $\Psi_{m}(p)$ can be interpreted as a momentum-spin component
probability amplitude just as in the nonrelativistic theory. The normalization
condition is
\begin{equation}
\int d^{3}p\sum_{m=\pm\frac{1}{2}}|\Psi_{m}(p)|^{2}=1.\label{eq:2.3}
\end{equation}
The expectations of the four-momentum operator and the rest-frame
spin $z$-component operator are, respectively,
\begin{equation}
\langle\,\psi\,|\,p^{\mu}\,|\,\psi\,\rangle=\int d^{3}p\sum_{m=\pm\frac{1}{2}}|\Psi_{m}(p)|^{2}\,p^{\mu}\quad\mathrm{and}\quad\langle\,\psi\,|\,s_{z}^{\circ}\,|\,\psi\,\rangle=\int d^{3}p\sum_{m=\pm\frac{1}{2}}|\Psi_{m}(p)|^{2}\,m.\label{eq:2.4}
\end{equation}

This is a \textit{covariant}, but not \textit{manifestly covariant}
theory. That means that the Lorentz transformation properties of the
expressions we write require a nontrivial derivation. In contrast,
for a manifestly covariant formalism, the transformation properties
of expressions are usually immediately obvious to the reader as those
expressions are written in terms of tensors and objects with simple,
known, transformation properties. The postulates of special relativity
do not require that all quantities of physical interest transform
as scalars, four-vectors and tensors. In combination with the rules
of quantum mechanics, they merely require that such transformations
be well-defined and unitary (or antiunitary in the case of time reversal)
and depend only on the translation, rotation or boost parameters.

The unitary transformations of the $\Psi_{m}(p)$ can be easily derived
\cite{Hoffmann2018d}. The technique is to apply the unitary (or antiunitary)
transformation to $|\,\psi\,\rangle$ and thus to the basis vectors,
then manipulate the expression into the form
\begin{equation}
U/A\,|\,\psi\,\rangle=\int\frac{d^{3}p}{\sqrt{\omega}}\sum_{m=\pm\frac{1}{2}}|\,p,s,m\,\rangle\Psi_{m}^{\prime}(p),\label{eq:2.5}
\end{equation}
then extract the $\Psi_{m}^{\prime}(p)$ by orthonormality.

The transformation rules for the Poincaré transformations are:
\begin{eqnarray}
\mathrm{Spacetime\ translations:}\quad\Psi_{m}^{\prime}(p) & = & \Psi_{m}(p)\,e^{+ip\cdot a},\nonumber \\
\mathrm{Rotations:}\quad\Psi_{m}^{\prime}(p) & = & \sum_{m^{\prime}=\pm\frac{1}{2}}\mathcal{D}_{mm^{\prime}}^{(\frac{1}{2})}(R)\,\Psi_{m^{\prime}}(R^{-1}p),\nonumber \\
\mathrm{Boosts}:\quad\Psi_{m}^{\prime}(p) & = & \sqrt{\gamma_{0}(1-\boldsymbol{\beta}_{0}\cdot\boldsymbol{\beta})}\sum_{m^{\prime}=\pm\frac{1}{2}}\mathcal{W}_{mm^{\prime}}^{(\frac{1}{2})}(p\leftarrow\Lambda^{-1}p)\,\Psi_{m^{\prime}}(\Lambda^{-1}p),\label{eq:2.6}
\end{eqnarray}
where $\boldsymbol{\beta}_{0}$ is the boost velocity, $\gamma_{0}=1/\sqrt{1-\boldsymbol{\beta}_{0}^{2}}$
and $\boldsymbol{\beta}=\boldsymbol{p}/\omega$ is the velocity of
the particle. For the inversions, we have
\begin{eqnarray}
\mathrm{Space\ inversion:}\quad\Psi_{m}^{\prime}(\omega,\boldsymbol{p}) & = & \eta\,\Psi_{m}(\omega,-\boldsymbol{p}),\nonumber \\
\mathrm{Time\ reversal:}\quad\Psi_{m}^{\prime}(\omega,\boldsymbol{p}) & = & (-)^{\frac{1}{2}+m}\,\Psi_{-m}^{*}(\omega,-\boldsymbol{p}).\label{eq:2.7}
\end{eqnarray}
In these expressions $\mathcal{W}$ is a matrix element of a Wigner
rotation, which can be evaluated by
\begin{equation}
W(\Lambda p\leftarrow p)=\Lambda^{-1}[\Lambda p]\,\Lambda\,\Lambda[p].\label{eq:2.8}
\end{equation}
where
\begin{equation}
\Lambda[p]\equiv\Lambda(\frac{\boldsymbol{p}}{\omega})\label{eq:2.9}
\end{equation}
and $\Lambda(\boldsymbol{\beta})$ is a function of the boost velocity,
$\boldsymbol{\beta}.$ Explicit forms of the Wigner rotations are
given in \cite{Halpern1968}. Two successive, noncollinear, boosts
(from the rest momentum to $p$ and then from $p$ to $\Lambda p$)
produce a boost (from the rest momentum to $\Lambda p$) preceded
by a rotation in the rest frame. This is the physics behind the Thomas
precession \cite{Jackson1975}.

These transformations are all unitary (antiunitary for time reversal)
in that they preserve the modulus-squared of the scalar product
\begin{equation}
|\langle\,\varphi\,|\,\psi\,\rangle|^{2}=\left|\int d^{3}p\sum_{m=\pm\frac{1}{2}}\varphi^{*}(p,m)\psi(p,m)\right|^{2}.\label{eq:2.10}
\end{equation}

The improper state vector for an electron or positron localized at
position $\boldsymbol{x}$ at time $t$ with spin $z$-component $m$
is given by Newton and Wigner \cite{Newton1949} (with $x^{\mu}=(t,\boldsymbol{x})^{\mu})$
as
\begin{equation}
|\,x,m\,\rangle=\int\frac{d^{3}p}{\sqrt{\omega}}|\,p,m\,\rangle\frac{e^{ip\cdot x}}{(2\pi)^{\frac{3}{2}}}.\label{eq:2.11}
\end{equation}
At equal times, these satisfy the orthogonality (and normalization)
condition
\begin{equation}
\langle\,t,\boldsymbol{x}_{1},m_{1}\,|\,t,\boldsymbol{x}_{2},m_{2}\,\rangle=\delta_{m_{1}m_{2}}\delta^{3}(\boldsymbol{x}_{1}-\boldsymbol{x}_{2}).\label{eq:2.12}
\end{equation}
Note that the spin component in each rest frame becomes the spin component
at a position and time.

The amplitudes for our state vector on this basis are then
\begin{equation}
\psi_{m}(x)\equiv\langle\,x,m\,|\,\psi\,\rangle=\int\frac{d^{3}p}{(2\pi)^{\frac{3}{2}}}\Psi_{m}(p)\,e^{-ip\cdot x},\label{eq:2.13}
\end{equation}
a Fourier transform of the momentum-spin component amplitudes as in
the nonrelativistic theory, with the form of the energy changed. The
unitary/antiunitary transformations of these position-spin component
amplitudes are given in \cite{Hoffmann2018d}. We merely note that
a boost transformation involves a nonlocal transformation of the amplitudes.

Now the total position probability density operator, summed over spin
components, is the projector
\begin{equation}
\rho(x)=\sum_{m=\pm\frac{1}{2}}|\,x,m\,\rangle\langle\,x,m\,|.\label{eq:2.14}
\end{equation}
We want to investigate whether this operator can be the zero component
of a four-component probability current density that transforms locally
as a four-vector function of the spacetime coordinates. A possible
obstacle to this is the fact that $\rho(x)$, unitarily transformed
on its own, transforms nonlocally.

For comparison, the Dirac charge density with a factor of the charge
removed can be written
\begin{equation}
\rho_{\mathrm{D}}(x)=\frac{m_{e}}{(2\pi)^{3}}\sum_{a=1}^{4}|\,x,a\,\rangle\langle\,x,a\,|,\label{eq:2.15}
\end{equation}
with
\begin{equation}
|\,x,a\,\rangle\equiv\int\frac{d^{3}p}{\omega}\sum_{m=\pm\frac{1}{2}}|\,p,m\,\rangle u_{ma}(p)\,e^{ip\cdot x}\label{eq:2.16}
\end{equation}
and
\begin{equation}
u_{ma}^{*}(p)=u_{am}(p)\label{eq:2.17}
\end{equation}
and
\begin{equation}
\begin{pmatrix}u_{1m}(p)\\
u_{2m}(p)\\
u_{3m}(p)\\
u_{4m}(p)
\end{pmatrix}=u(p,m).\label{eq:2.18}
\end{equation}
The $|\,x,a\,\rangle$ boost locally.

\section{The commutators between the boost generators and the components of
a four-vector}

Under the Lorentz transformations as defined, the spacetime origin
is invariant, so we need only attempt to construct $J^{\mu}(0),$
which is then required to transform as a four-vector. Then we may
translate it to general $x$ using Eq.~(\ref{eq:1.2}).

For a boost by infinitesimal rapidity, $\boldsymbol{\boldsymbol{\zeta},}$
Eq. (\ref{eq:1.1}) becomes
\begin{align}
(1+i\boldsymbol{\zeta}\cdot\boldsymbol{K})\,J^{0}(0)\,(1-i\boldsymbol{\zeta}\cdot\boldsymbol{K}) & =J^{0}(0)+\boldsymbol{\zeta}\cdot\boldsymbol{J}(0),\nonumber \\
(1+i\boldsymbol{\zeta}\cdot\boldsymbol{K})\,\boldsymbol{J}(0)\,(1-i\boldsymbol{\zeta}\cdot\boldsymbol{K}) & =\boldsymbol{J}(0)+J^{0}(0)\,\boldsymbol{\zeta}.\label{eq:3.1}
\end{align}
This gives the two sets of commutators
\begin{align}
i[\boldsymbol{K},J^{0}(0)] & =\boldsymbol{J}(0)\label{eq:3.2}
\end{align}
and
\begin{equation}
i[K_{i},J_{j}(0)]=\delta_{ij}J^{0}(0).\label{eq:3.3}
\end{equation}

We need the representation of the boost generators, $\boldsymbol{K},$
acting on the $\Psi_{m}(p).$ We take
\begin{equation}
(1-i\boldsymbol{\zeta}\cdot\boldsymbol{K})\,|\,\psi\,\rangle=\int\frac{d^{3}p}{\sqrt{\omega}}\sum_{m=\pm\frac{1}{2}}|\,p,s,m\,\rangle\Psi_{m}^{\prime}(p),\label{eq:3.4}
\end{equation}
with $\Psi_{m}^{\prime}(p)$ given by the third equation of the set
Eq. (\ref{eq:2.6}), approximated to first order in $\boldsymbol{\zeta}.$
We need to approximate the explicit formula for the Wigner rotation.
We find
\begin{equation}
\boldsymbol{K}\,|\,\psi\,\rangle=\int\frac{d^{3}p}{\sqrt{\omega}}\sum_{m=\pm\frac{1}{2}}|\,p,s,m\,\rangle\{-\frac{i}{2}\{\omega,\frac{\partial}{\partial\boldsymbol{p}}\}+\frac{1}{2}\frac{\boldsymbol{\sigma}\times\boldsymbol{p}}{\omega+m_{e}}\}\Psi(p),\label{eq:3.5}
\end{equation}
where the $\boldsymbol{\sigma}$ are the usual Pauli matrices. We
note that this representation of the operator is explicitly Hermitian,
as it must be since the form of the scalar product is simply given
by Eq. (\ref{eq:2.10}). The anticommutator of two operators is defined
by $\{A,B\}=AB+BA.$

\section{Construction and test of a possible four-vector probability current
density}

First we write $J^{0}(0)=\rho(x)$ in momentum space using Eqs. (\ref{eq:2.11})
and (\ref{eq:2.14}). This gives
\begin{equation}
J^{0}(0)=\frac{1}{(2\pi)^{3}}\int\frac{d^{3}p_{a}}{\sqrt{\omega_{a}}}\int\frac{d^{3}p_{b}}{\sqrt{\omega_{b}}}\sum_{m=\pm\frac{1}{2}}|\,p_{a},m\,\rangle\langle\,p_{b},m\,|.\label{eq:4.1}
\end{equation}
From Eq. (\ref{eq:3.2}), using Eq. (\ref{eq:3.5}), this gives the
spatial part of the possible four-vector as
\begin{equation}
\boldsymbol{J}(0)=\frac{1}{(2\pi)^{3}}\int\frac{d^{3}p_{a}}{\sqrt{\omega_{a}}}\int\frac{d^{3}p_{b}}{\sqrt{\omega_{b}}}\sum_{m=\pm\frac{1}{2}}|\,p_{a},m_{a}\,\rangle\{\frac{1}{2}(\boldsymbol{\beta}_{a}+\boldsymbol{\beta}_{b})\delta_{m_{a}m_{b}}+\frac{i}{2}[\frac{\boldsymbol{\sigma}\times\boldsymbol{p}_{a}}{\omega_{a}+m_{e}}-\frac{\boldsymbol{\sigma}\times\boldsymbol{p}_{b}}{\omega_{b}+m_{e}}]_{m_{a}m_{b}}\}\langle\,p_{b},m_{b}\,|,\label{eq:4.2}
\end{equation}
where $\boldsymbol{\beta}_{a/b}=\boldsymbol{p}_{a/b}/\omega_{a/b}.$
We note the explicit Hermiticity of this expression.

Now we must test the other commutators, Eq. (\ref{eq:3.3}). To simplify
the calculation, we test
\begin{equation}
i\sum_{i=1}^{3}[K_{i},J_{i}(0)]=3\,J^{0}(0),\label{eq:4.3}
\end{equation}
which is a necessary but not sufficient condition for covariance.

After some calculation, we find
\begin{equation}
i\sum_{i=1}^{3}[K_{i},J_{i}(0)]=\frac{1}{(2\pi)^{3}}\int\frac{d^{3}p_{a}}{\sqrt{\omega_{a}}}\int\frac{d^{3}p_{b}}{\sqrt{\omega_{b}}}\sum_{m=\pm\frac{1}{2}}|\,p_{a},m\,\rangle\{3-\frac{1}{4}|\boldsymbol{\beta}_{a}-\boldsymbol{\beta}_{b}|^{2}-\frac{1}{2}\left|\frac{\boldsymbol{p}_{a}}{\omega_{a}+m_{e}}-\frac{\boldsymbol{p}_{b}}{\omega_{b}+m_{e}}\right|^{2}\}\langle\,p_{b},m\,|,\label{eq:4.4}
\end{equation}
Hermitian and rotationally invariant, as expected. Clearly the condition
Eq. (\ref{eq:4.4}) is not satisfied, so it is not possible to construct
a four-vector probability current for the electron or positron.

\section{The spinless case, comparison with other work and the case of general
spin}

It is a simple matter to eliminate spin from these equations and arrive
at the candidate four components for the spinless case:
\begin{equation}
J_{0}^{0}(0)=\frac{1}{(2\pi)^{3}}\int\frac{d^{3}p_{a}}{\sqrt{\omega_{a}}}\int\frac{d^{3}p_{b}}{\sqrt{\omega_{b}}}\,|\,p_{a}\,\rangle\langle\,p_{b}\,|\label{eq:5.1}
\end{equation}
and
\begin{equation}
\boldsymbol{J}_{0}(0)=\frac{1}{(2\pi)^{3}}\int\frac{d^{3}p_{a}}{\sqrt{\omega_{a}}}\int\frac{d^{3}p_{b}}{\sqrt{\omega_{b}}}\,|\,p_{a}\,\rangle\frac{1}{2}(\boldsymbol{\beta}_{a}+\boldsymbol{\beta}_{b})\langle\,p_{b}\,|.\label{eq:5.2}
\end{equation}
This was the form obtained by Rosenstein and Horwitz \cite{Rosenstein1985}.
The commutator of Eq. (\ref{eq:4.3}) comes out in this case to be
\begin{equation}
i\sum_{i=1}^{3}[K_{i},J_{0i}(0)]=\frac{1}{(2\pi)^{3}}\int\frac{d^{3}p_{a}}{\sqrt{\omega_{a}}}\int\frac{d^{3}p_{b}}{\sqrt{\omega_{b}}}\,|\,p_{a}\,\rangle\{3-\frac{1}{4}|\boldsymbol{\beta}_{a}-\boldsymbol{\beta}_{b}|^{2}\}\langle\,p_{b}\,|,\label{eq:5.3}
\end{equation}
not equal to $3\,J_{0}^{0}(0).$ So this proposed current is not covariant.

It is informative to try to write the candidate spinless four-current
in manifestly covariant form, using
\begin{equation}
\frac{1}{2}(1,\boldsymbol{\beta}_{a/b})^{\mu}=\frac{1}{2}\frac{p_{a/b}^{\mu}}{\omega_{a/b}}.\label{eq:5.4}
\end{equation}
This gives
\begin{equation}
J_{0}^{\mu}(0)=\frac{1}{(2\pi)^{3}}\int\frac{d^{3}p_{a}}{\omega_{a}}\int\frac{d^{3}p_{b}}{\omega_{b}}\,|\,p_{a}\,\rangle\frac{1}{2}\{\sqrt{\frac{\omega_{b}}{\omega_{a}}}\,p_{a}^{\mu}+\sqrt{\frac{\omega_{a}}{\omega_{b}}}\,p_{b}^{\mu}\}\langle\,p_{b}\,|,\label{eq:5.5}
\end{equation}
clearly not covariant.

We note that Kowalski and Rembieli\'{n}ski \cite{Kowalski2011} derive
a four-component object, with the position probability density (Eq.
(\ref{eq:2.14})) as the zero component, that is locally conserved
in one frame. This is done by solving the local conservation equation.
This object clearly does not have the required transformation properties.

For general spin $s=0,\frac{1}{2},1,\frac{3}{2},\dots,$ we start
with the basis vectors $|\,p,s,m\,\rangle\ (m=-s,-s+1,\dots,s-1,s).$
Much of the calculation is similar to the two particular cases just
considered. The Pauli matrices are replaced by
\begin{equation}
\frac{1}{2}[\boldsymbol{\sigma}]_{m_{a}m_{b}}\rightarrow\langle\,s,m_{a}\,|\,\boldsymbol{J}\,|\,s,m_{b}\,\rangle.\label{eq:5.6}
\end{equation}
It is clear that the term $-\frac{1}{4}|\boldsymbol{\beta}_{a}-\boldsymbol{\beta}_{b}|^{2}$
will always appear within the representations of the commutator Eq.
(\ref{eq:4.3}), along with rotationally invariant terms contributed
by the spin, destroying any possibility of covariance.

\section{Conclusions}

We have seen that the relativistic probability density for an electron/positron
cannot be incorporated as the zero component of a four-vector probability
current density. The same is true for any spin.

The time evolution of an expectation value of the position probability
density operator can be calculated by taking an expectation of Eq.
(\ref{eq:2.14}) or (\ref{eq:4.1}). The result is relativistic since
we know exactly how any boosted frame would view this evolution, and
the integral over all space is invariant. We have found that this
density does not ``flow'' like the density of a relativistic fluid
with a local conservation law. This result adds to the discussion
around the results of Hegerfeldt \cite{Hegerfeldt1974,Hegerfeldt1980,Hegerfeldt1985}
regarding violation of a classical notion of causality in relativistic
quantum mechanics. Since the probability density does not flow like
the density of a fluid, it certainly does not flow like the density
of a fluid with a field of velocities that are constrained to be less
than the speed of light in magnitude.

This result strongly supports the conclusion that the Dirac current
is the unique choice for the electromagnetic current of an electron
or positron (with the caveat that a method must be found to modify
the g-factor while preserving the positive-semidefiniteness of the
charge density, a desirable constraint).

Another aim of this paper is to promote the use of relativistic probability
amplitudes. It would make little sense to second quantize these amplitudes
to form nonlocally transforming field operators. Yet these amplitudes
have a definite role to play in quantum field theory. If one is to
construct a realistic description of a scattering experiment, one
needs to describe the essentially free initial and final states with
wavepackets, with the probability distributions in momentum, position
and spin specified. To do this it is essential to use relativistic
probability amplitudes.

In addition, it has been shown that the use of wavepackets eliminates
some divergences in scattering calculations \cite{Hoffmann2017a}.

\bibliographystyle{apsrev4-1}

\end{document}